\documentclass[reprint,superscriptaddress,nofootinbib,amsmath,amssymb,aps,prl,onecolumn,notitlepage]{revtex4-1}

\usepackage[a4paper, margin=2.50cm]{geometry}

\usepackage[utf8]{inputenc}
\usepackage[T1]{fontenc}
\usepackage[english]{babel}
\usepackage{amsthm}
\newtheorem{theorem}{Theorem}[section]
\newtheorem{corollary}{Corollary}[theorem]

\newtheorem{example}[theorem]{Example}
\newtheorem{definition}[theorem]{Definition} 

\usepackage{xcolor} 
\usepackage{amsmath, bm, amsfonts, mathtools, amssymb, amsthm, slashed, graphicx, setspace}
\usepackage{multirow}
\DeclareMathOperator{\tr}{Tr} 
\usepackage{fixmath}

\usepackage{booktabs}
\usepackage{subfigure}
\usepackage{verbatim}

\usepackage{graphicx}
\usepackage{bbm}
\usepackage{graphicx}
\usepackage{fancyhdr}
\usepackage{hyperref}  
\usepackage{bm}

\usepackage{enumitem} 

\usepackage{dsfont, natbib}

\usepackage{etoolbox}
\makeatletter
\patchcmd{\@maketitle}{\newpage}{}{}{} 
\makeatother

\def\>{\rangle}
\def\<{\langle}

\makeatletter

\DeclarePairedDelimiterX\norm[1]{\lVert}{\rVert}{{#1}}

\DeclarePairedDelimiterX\ket[1]{\lvert}{\rangle}{{#1}}
\DeclarePairedDelimiterX\bra[1]{\langle}{\rvert}{{#1}}
\DeclarePairedDelimiterX\braket[2]{\langle}{\rangle}{%
  {#1}\hspace*{0.2ex}\delimsize\vert\hspace*{0.2ex}{#2}%
}
\DeclarePairedDelimiterX\ketbra[2]{\lvert}{\rvert}{%
  {#1}\delimsize\rangle\hspace*{-0.25ex}\delimsize\langle{#2}%
}
\DeclarePairedDelimiterX\proj[1]{\lvert}{\rvert}{%
  {#1}\delimsize\rangle\hspace*{-0.25ex}\delimsize\langle{#1}%
}
\DeclarePairedDelimiterX\matrixel[3]{\langle}{\rangle}{%
  {#1}\hspace*{0.2ex}\delimsize\vert\hspace*{0.2ex}{#2}%
  \hspace*{0.2ex}\delimsize\vert\hspace*{0.2ex}{#3}%
}
\DeclarePairedDelimiterX\dmatrixel[2]{\langle}{\rangle}{%
  {#1}\hspace*{0.2ex}\delimsize\vert\hspace*{0.2ex}{#2}%
  \hspace*{0.2ex}\delimsize\vert\hspace*{0.2ex}{#1}%
}
\DeclarePairedDelimiterX\innerprod[2]{\langle}{\rangle}{%
  {#1},\hspace*{0.2ex}{#2}%
}

\def\DD{D\DD@}
\DeclarePairedDelimiterX\DD@[2]{(}{)}{%
  {#1}\delimsize\Vert{#2}%
}

\newcommand\etal{\emph{et al}\@ifnextchar.\relax{\emph{.\@}}}

\colorlet{phfcolor}{red!50!orange}
\colorlet{phfrmcolor}{phfcolor!50!gray!35!white}
\def\phf{\@ifnextchar[\phf@\phf@@}
\newcommand\phf@[1][]{{\color{phfcolor}\fontfamily{cmbr}\fontseries{sb}\selectfont \;[\,{#1}\,]\;}}
\newcommand\phf@@[1]{{\color{phfcolor}{#1}}}

\makeatother

\begin{document}

\fancyhead[C]{\sc \color[rgb]{0.4,0.2,0.9}{Quantum Thermodynamics book}}
\fancyhead[R]{}

\title{Quantum thermodynamics with multiple conserved quantities}

\author{Erick Hinds-Mingo}
\affiliation{QOLS, Blackett Laboratory, Imperial College London, London SW7 2AZ, United Kingdom}

\author{Yelena Guryanova}
\affiliation{Institute for Quantum Optics and Quantum Information (IQOQI), Boltzmanngasse 3 1090, Vienna, Austria}

\author{Philippe Faist}
\affiliation{Institute for Quantum Information and Matter, Caltech, Pasadena CA, 91125 USA}

\author{David Jennings}
\affiliation{Department of Physics, University of Oxford, Oxford, OX1 3PU, United Kingdom}
\affiliation{QOLS, Blackett Laboratory, Imperial College London, London SW7 2AZ, United Kingdom}

\date{\today}

\begin{abstract}

In this chapter we address the topic of quantum thermodynamics in the presence of additional observables beyond the energy of the system. In particular we discuss the special role that the generalized Gibbs ensemble plays in this theory, and derive this state from the perspectives of a micro-canonical ensemble, dynamical typicality and a resource-theory formulation. A notable obstacle occurs when some of the observables do not commute, and so it is impossible for the observables to simultaneously take on sharp microscopic values. We show how this can be circumvented, discuss information-theoretic aspects of the setting,  and explain how thermodynamic costs can be traded between the different observables. Finally, we discuss open problems and future directions for the topic.

\end{abstract}

\maketitle
\thispagestyle{fancy}


\section{Introduction}

Thermodynamics has been remarkable in its applicability to a vast array of systems.
Indeed, the laws of macroscopic thermodynamics have been successfully applied to the studies of magnetization~\cite{Magnetization1,Magnetization2}, superconductivity~\cite{Superconductivity}, cosmology~\cite{Cosmology2}, chemical reactions~\cite{Chemistry} and biological phenomena~\cite{WhatIsLife,Biology}, to name a few fields. In thermodynamics, energy plays a key role as a thermodynamic potential, that is, as a function of the other thermodynamic variables which characterizes all the thermodynamic properties of the system.  In the presence of thermodynamic reservoirs, however, physical quantities which are globally conserved can be exchanged with the reservoirs, and the thermodynamic properties of the system are more conveniently expressed in terms of other potentials.  For example, in the grand canonical ensemble, particle number $N$ as well as energy $E$ are exchanged with a reservoir. The relevant thermodynamic potential becomes,
     \begin{align}
       F = E  - \mu N   - T S\ ,
     \end{align}
where $S$ is the entropy, and where $T$ and $\mu$ are the \emph{temperature} of the heat bath and the \emph{chemical potential} of the particle reservoir, respectively.
The chemical potential $\mu$ acts as an `exchange rate' between particle number and energy, in the same way that temperature $T$ acts as an exchange rate between entropy and energy.  Thus, $\mu$ describes the energetic cost of adding another particle to a gas at constant entropy. In classical equilibrium thermodynamics, for $k$ conserved quantities, or \textit{charges}\footnote{terms we will use interchangeably} $\{Q_k\}$ the function $U(S,Q_1,...,Q_r)$ characterises the internal energy of the system. Each charge has an associated `exchange rate' $\mu_i:= \frac{\partial U}{\partial Q_i}$ that governs the response in energy when one varies the equilibrium value of $Q_i$. 

Equivalently, one can use the entropic formulation of thermodynamics to interpret the change in a system at equilibrium. The `entropic response' to any change in an extensive variable is given by the generalised temperatures $\beta_k := \frac{\partial S}{\partial Q_k}$, which are related to the chemical potentials by $\beta_k = \frac{\mu_k}{T}$.

A canonical example is provided by a macroscopic system with angular momentum observables $(J_x, J_y,J_z) =: \boldsymbol{J}$, which has a non-zero polarization in these observables along some axis when in thermodynamic equilibrium. The internal energy for this thermodynamic system is therefore a function $U(S, J)$ where $J = \boldsymbol{J}\cdot \boldsymbol{n}$, and $\boldsymbol{n}$ is a unit vector along the distinguished axis. Any relaxation to equilibrium occurs through the microscopic exchange of angular momentum with the environment. Indeed, the environment must play a role of defining a thermodynamic constraints for the system, in both energy and angular momentum. Subject to these constraints, the equilibrium thermodynamics is determined by maximization of the entropy $S$ in the usual way.

More recently, researchers have been experimenting with the idea that quantum mechanics could also present interesting and novel features in conjunction with thermodynamic systems on the microscopic scale. While a well-established framework exists for macroscopic equilibrium systems, it is less clear how to handle finite-sized systems with multiple conserved charges that may display quantum mechanical features, such as complementarity.

A straightforward approach to studying thermodynamics of system possessing multiple conserved charges is to apply the quantum version of Jaynes' principle~\cite{Jaynes1,Jaynes2}, which prescribes that one should maximize the (von Neumann) entropy subject to constraints on the average values $\{ v_j \}$ of the conserved charges $\{ Q_j \}$:
  \begin{equation}
  \begin{aligned}
    \mbox{maximize:} &\quad S(\tau) \\
    \mbox{subject to:} &\quad \tr(Q_j\tau) = v_j\ \ \forall\ j\ .
  \end{aligned}
\end{equation}
The unique solution is the so-called \emph{generalized Gibbs state} (or \emph{generalized Gibbs ensemble})
\begin{align}
  \label{eq:GGS}
  \tau = \frac{e^{-(\beta_1 Q_1 + \cdots + \beta_k Q_k)} }{\mathcal{Z}} 
  \,,
\end{align}
where the $\{ \beta_j \}$ are generalized inverse temperatures or generalized chemical potentials which are determined by the $\{ v_j \}$.  The partition function $\mathcal{Z} = \tr [e^{-(\beta_1 Q_1 + \cdots + \beta_k Q_k)}]$ normalises the state.

While the generalized Gibbs state is distinguished as resulting from Jaynes' principle, it is not clear how this state may be interpreted as the thermal state of the system on a physical level. 

As an illustration, we could consider a single spin-1/2 particle with degenerate Hamiltonian $H=0$ and spin angular momentum operators given by the Pauli operators $(\sigma_x, \sigma_y, \sigma_z)$. We might wish to create constraints on this system such that the expectation values are fixed $\<\sigma_x\> =s_x$ and $\<\sigma_y\>=s_y$ for constant $s_x,s_y$. In accordance with the maximum entropy formulation, the state we infer from these constraints admits the form
\begin{equation}
\tau = \frac{e^{-\beta_x \sigma_x - \beta_y \sigma_y}}{\mathcal{Z}},
\end{equation}
for $\mathcal{Z} = \tr[e^{-\beta_x \sigma_x - \beta_y \sigma_y}]$, and some constants $\beta_x$ and $\beta_y$. In realisations of this setup, one might expect the quantum system could interact with its environment and relax under some dynamics to the generalized Gibbs state, which is addressed in \cite{EquilibrationChap}  $\ $when dealing with dynamical equilibration.

Subtleties arise in making this assumption. The existence of a physical map that realises this state subject to the constraints turns out to be forbidden by quantum mechanics. Specifically it is known the so-called `pancake map' that projects a spin-1/2 state onto the disk in $X-Y$ plane would be allowed if quantum mechanics had no entanglement, but in the presence of entanglement creates negative probabilities and is thus an unphysical transformation (technically, this map is not completely positive)~\cite{NonCommutativity}. 

Given this and other subtleties (such as even defining a micro-canonical ensemble for non-commuting charges), a basic question for contemporary quantum thermodynamics is therefore to understand whether the thermal state of a quantum system is indeed given by~\eqref{eq:GGS} in the presence of multiple, possibly non-commuting charges~\cite{Halpern}.

  In the situation where the charges commute, it was already shown that Landauer erasure could be carried out by utilizing physical quantities other than energy~\cite{Varcoe,vaccaro2011information,InformationErasureChap}, and that the resource theory approach to quantum thermodynamics~\cite{Brandao2013,Brandao2015PNAS_secondlaws} could be generalized to multiple physical quantities~\cite{YungerHalpern2016PRE_beyond,Halpern,Weilenmann2016PRL_axiomatic}.
  
  On the other hand, the generalized Gibbs ensemble was given considerable interest in the context of systems which are integrable, i.e., which do not thermalize, as further constants of motion constrain the evolution of the system~\cite{rigol2007relaxation,gogolin2016equilibration}.  Such situations have been demonstrated experimentally~\cite{Cassidy2011PRL_integrable,Langen2015Sci_GGE}.
Work extraction was also studied in the context of generalized Gibbs ensembles, bridging both aspects~\cite{perarnau2016work}.

\begin{figure}
    \centering
    \includegraphics[width=120mm]{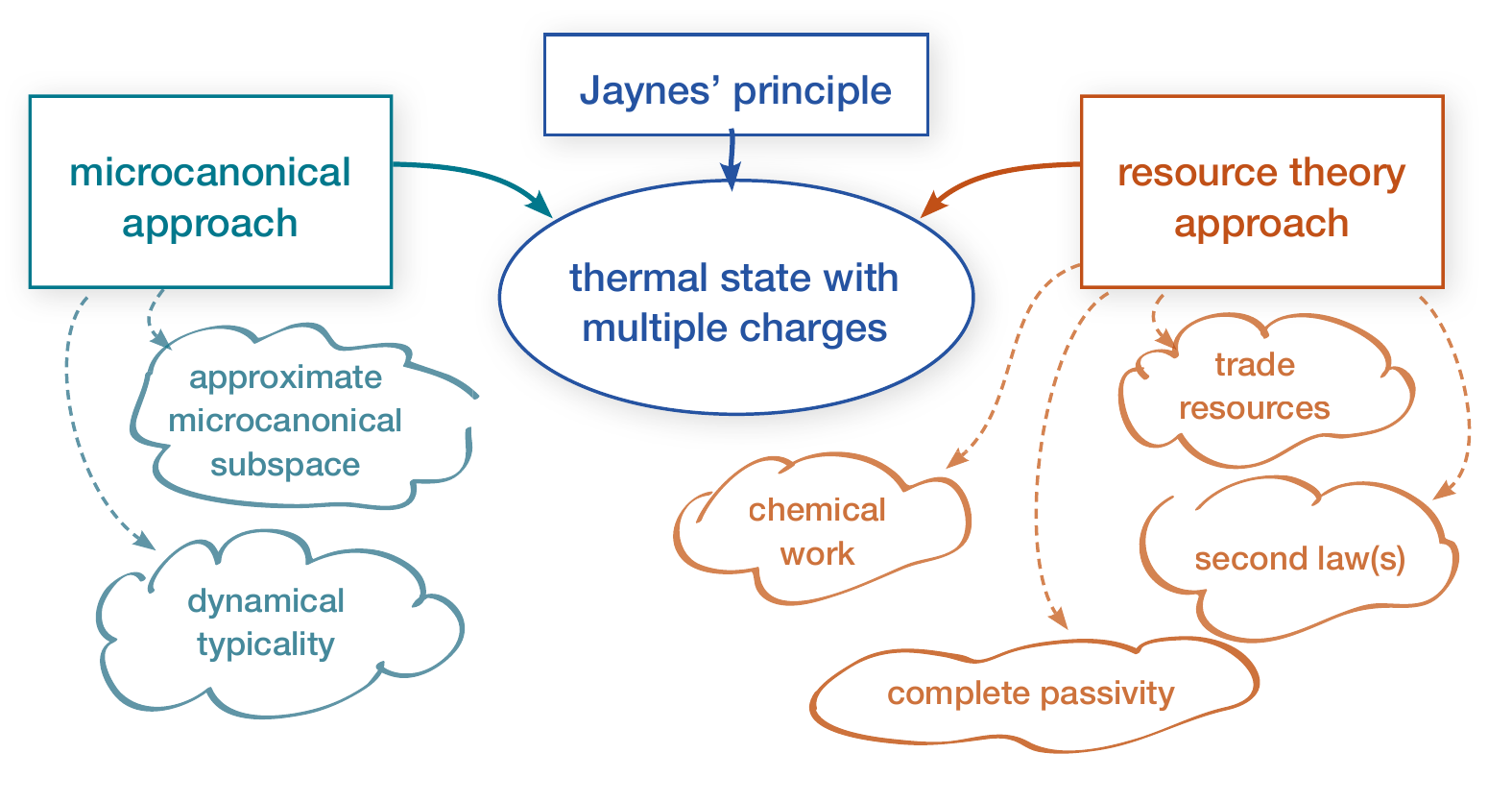}
    \caption{The thermal state in the presence of multiple reservoirs corresponding to different physical charges can be determined via several approaches.  If the charges commute, the equilibrium state of a system is a corresponding grand-canonical state, which is readily derived by either considering a microcanonical ensemble over the system and the reservoirs, by Jaynes' principle, or via resource-theoretic considerations~\cite{YungerHalpern2016PRE_beyond}.  While it is straightforward to apply Jaynes' principle to the case of non-commuting charges, the other two approaches need to be adapted.  This chapter reviews how to generalize these approaches to non-commuting charges, introducing along the way the notions of an approximate microcanonical subspace and how it connects with dynamical typicality, as well as the idea of trading resources using batteries, the second law and complete passivity for non-commuting charges, and chemical work.}
    \label{fig:OverviewApproaches}
  \end{figure}

In this chapter, we derive the form of this state through the lens of two different approaches introduced in refs.~\cite{MicroCanonical,MultipleConserved, NonCommutativity}.  In the process, we explain how these approaches fit together with the usual concepts of statistical mechanics such as the microcanonical state on one hand, and with the second law of thermodynamics and complete passivity on the other (\autoref{fig:OverviewApproaches}). We primarily address the topic using tools from quantum information theory, which provide novel approaches for dynamical typicality and equilibration theory of arbitrary charges $\{Q_1, \dots ,Q_n\}$ that may have non-trivial commutation relations among themselves.  In addition, the recent resource-theoretic approach to thermodynamics has allowed a well-defined framework in which to analyse quantum thermodynamics without the need to use notions of `heat' or `work' as defining concepts. We also discuss how multiple conserved charges can fit into such an approach and discuss the subtleties that can arise when one attempts to do so. Finally we discuss information-theoretic aspects of quantum thermodynamics with multiple conserved charges and provide a generalized Landauer bound that shows that erasure can be carried out at no energetic cost. Lastly, we discuss the status of this topic within contemporary quantum thermodynamics and the core challenges that exist going forward.

\section{Microcanonical approach}

In this section we present an approach for deriving the form of the generalized Gibbs state by generalizing the concept of a microcanonical subspace to noncommuting charges.

In the presence of reservoirs exchanging commuting charges, the thermal state of the system can be derived by considering the system together with the reservoirs as a huge system in a microcanonical state, i.e., the maximally mixed state living in the common subspace of fixed total value of each charge, and then tracing out the reservoirs. For instance, if a system $s$ is in contact with both a heat reservoir $R_1$ and a particle reservoir $R_2$, we assume that the total system has fixed values of energy $E$ and number of particles $N$ and we consider the common eigenspace of the Hamiltonian and the number operator on the total system, the projector on which we denote by $\Pi^{(E,N)}$.  Due to the fundamental postulate of statistical mechanics, the corresponding microcanonical state is $\Omega^{(E,N)} = \Pi^{(E,N)}/\tr[\Pi^{(E,N)}]$, and under mild assumptions one can show that for large reservoirs the reduced state on the system is the grand canonical ensemble, $\tau = \tr_{R_1R_2}[\Omega^{(E,N)}] \approx e^{-\beta({H}_s - \mu{N}_s)}/\mathcal{Z}(\beta,\mu)$, where $\beta,\mu$ are the inverse temperature and the chemical potential, respectively, and where ${H}_s$ and ${N}_s$ are the Hamiltonian and number operator of the system  (see refs.~\cite{PhysRevLett.96.050403,10.1038/nphys444}).

\begin{figure}
      \centering
      \includegraphics{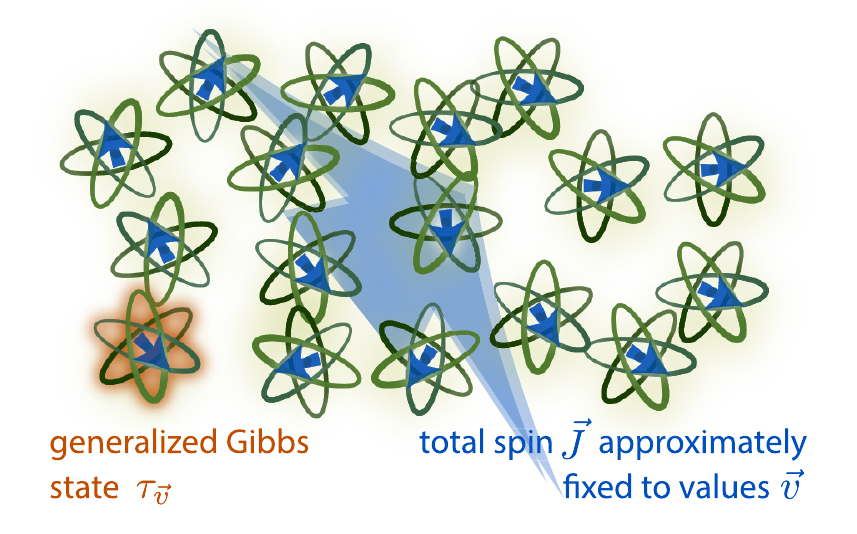}
      \caption{The thermal state of a system in the presence of multiple noncommuting conserved charges can be derived from an approximate microcanonical subspace on many copies of the system.  (For instance, the charges might be components of the spin $\vec{J}$.) This subspace has the property that states within the subspace have statistics for each charge that are sharply peaked around a given set of values $\vec{v}$.  If we trace out all systems except one, the reduced state is close to the generalized Gibbs state $\tau_{\vec v} \propto \exp(-\sum \beta_i J_i)$ for appropriate generalized chemical potentials $\{\beta_i\}$.
        Figure reproduced from ref.~\cite{MicroCanonical} (CC-BY) with adapted notation.}
      \label{fig:ApproxMicrocan}
    \end{figure}

If the charges do not commute, then there are no common eigenspaces for the different charges, and we cannot define $\Pi^{(E,N)}$ as above.  However, this approach can be adapted so that it applies to noncommuting charges.  The key idea, proposed by Yunger Halpern~\etal~\cite{MicroCanonical}, is the following: If we consider many copies of the system, there may be no exact common eigenspaces, but we may define instead an \emph{approximate microcanonical subspace}. Instead of fixing the values of the charges exactly, the approximate microcanonical subspace only fixes them approximately, by considering states which have sharply peaked statistics for each charge (\autoref{fig:ApproxMicrocan}).  So, we may consider the maximally mixed state supported on this subspace, and it turns out that tracing out the reservoirs yields a thermal state of the required form~\eqref{eq:GGS}.

The argument of ref.~\cite{MicroCanonical} goes as follows.  Consider a system $s$ with multiple physical charges represented by operators $Q_1,\ldots,Q_k$, which do not necessarily commute.  We consider $N$ copies of $s$; conceptually we might, for instance, think of the first copy as being the system of interest, and the rest as parts forming a large bath.  The composite average observables are defined as:
\begin{equation}
  \bar{Q}_i := \frac{1}{N} \sum_{l=0}^{N-1} \mathds{1}^{\otimes l} \otimes  Q_i \otimes \mathds{1}^{\otimes (N-1 - l)}.
\end{equation}
Because the charges do not commute, there may be no common eigenspaces to the set of operators $\bar{Q}_i$.  However, the noncommutativity of the $\bar{Q}_i$ ``wears out'' over many copies: One readily sees that $\norm{[\bar{Q}_i,\bar{Q}_j]} \sim 1/N \to 0$ as $N\to\infty$.  Intuitively, it should be possible to find a subspace which behaves approximately like a microcanonical subspace over the $N$ systems for large $N$.  For commuting charges, a defining property of the usual microcanonical subspace is that it contains all states which have a fixed given value for each charge.  So, for noncommuting charges, a natural loosening of this condition is to require that any state in the subspace has sharply peaked statistics for each charge, and conversely, that any state with sharply peaked statistics for each charge has large overlap with the subspace.  Such a subspace is called an \emph{approximate microcanonical subspace}:
\begin{definition}
  An \emph{approximate microcanonical subspace} $\mathcal{M}$ associated to fixed average values $\{ v_j \}$ of the charges $\{ Q_j \}$, is a subspace of $\mathcal{H}_s^{\otimes N}$ obeying the two following properties:
  \begin{enumerate}[label=(\roman*)]
  \item Any state $\rho$ with support inside $\mathcal{M}$ produces sharp statistics for measurements of $\bar{Q}_j$ for all $j$:
    \begin{align}
      \text{$\rho$ in $\mathcal{M}$} \qquad \Rightarrow \qquad
      \Pr\bigl[(\text{outcome of $\bar{Q}_j$}) \approx v_j\bigr] \approx 1\quad\forall~j\ ;
    \end{align}
  \item Conversely, any state $\rho$ producing sharp statistics for all $Q_j$ has high overlap with $\mathcal{M}$:
    \begin{align}
      \Pr\bigl[(\text{outcome of $\bar{Q}_j$}) \approx v_j\bigr] \approx 1 \quad \forall~j\
      \qquad \Rightarrow\qquad \tr[P\rho] \approx 1\ ,
    \end{align}
    where $P$ is the projector onto the subspace $\mathcal{M}$.
  \end{enumerate}
\end{definition}

(This definition is made technically precise by introducing an additive tolerance parameter for each approximation denoted by `$\approx$' above~\cite{MicroCanonical}.)  It is not obvious that this ``epsilonification'' of the usual microcanonical ensemble still has the properties we would like---in particular, that the reduced state on a single system is close to the generalized Gibbs state~\eqref{eq:GGS}.  It turns out, though, that any approximate microcanonical subspace has this property.  We can quantify the distance between the reduced states on each system and the generalized Gibbs state using the relative entropy $\DD{\rho}{\sigma} = \tr(\rho(\log\rho-\log\sigma))$.  The average relative entropy to the generalized Gibbs state of the reduced states on each system is small:
\begin{theorem}
  Let $\mathcal{M}$ be any approximate microcanonical subspace on $\mathcal{H}_s^{\otimes N}$, and define the approximate microcanonical state $\Omega = P/\tr[P]$, where $P$ is the projector onto the subspace $\mathcal{M}$.  Then, on average, the reduced state on system $\ell$ looks like the generalized Gibbs state:
  \begin{align}
    \frac1N\sum_{\ell=1}^N
    \DD*{\tr_{1,\ldots \ell-1,\ell+1\ldots, N}\bigl
    	[\Omega\bigr] 
    }{
    \frac{e^{-\sum_i \beta_i Q_i}}{\mathcal{Z}(\beta_1,\ldots,\beta_k)}
    } \sim \frac1{\sqrt N}\ ,
    \label{eq:approx-microcan-gives-reduced-thermal-state}
  \end{align}
  for appropriate generalized inverse temperatures $\{ \beta_j \}$.
\end{theorem}

It is left to show that it is possible to actually construct an approximate microcanonical subspace, i.e., that such a subspace actually exists for any collection of observables and for large enough $N$.
We may construct an approximate microcanonical subspace as follows.  A theorem by Ogata~\cite{Ogata} guarantees that there exist operators $\{\bar{Y}_j\}$ on the $N$ systems which are close to the $\{\bar{Q}_j\}$, and which \emph{do} commute exactly: $[\bar{Y}_i, \bar{Y}_j] = 0$ for all $i,j$ and $\norm{\bar{Y}_j - \bar{Q}_j}\to 0$ as $N\to\infty$.  So, given a set of charge values $\{ v_j \}$, it is possible to consider the common eigenspace $\mathcal{M}_\mathrm{com}$ of all the $\{ \bar{Y}_j \}$ corresponding to eigenvalues which are approximately equal to $\{ v_j \}$.  The subspace $\mathcal{M}_\mathrm{com}$ is an approximate microcanonical subspace for the commuting observables $\bar{Y}_j$ and values $\{ v_j \}$, because a usual microcanonical subspace is in particular an approximate microcanonical subspace.\footnote{In ref.~\cite{MicroCanonical}, the argument considers more generally an approximate microcanonical subspace over $m$ copies of the whole system, i.e., over a total of $Nm$ systems.}  Now, because $\bar{Y}_j\approx \bar{Q}_j$, a state with sharp statistics for $\bar{Y}_j$ also has sharp statistics for $\bar{Q}_j$ and \emph{vice versa}.  Hence finally, the subspace $\mathcal{M}_\mathrm{com}$ is in fact (after adapting the tolerance parameters) also an approximate microcanonical subspace for the $\{ \bar{Q}_j \}$.  The construction works for any $N$ large enough, and the tolerance parameters of the subspace may be taken to all go to zero simultaneously as $N\to\infty$.

While Ogata's theorem provides an intuitive way to construct an approximate microcanonical subspace, we note that other constructions are possible.  Furthermore it is possible to ensure that the subspace is manifestly permutation-invariant~\cite{AndreasUnpublishedNotes}.

\subsection{Dynamical typicality and evolution}

One may ask, is there any sense in which the system can be argued to \emph{evolve} towards the generalized Gibbs state~\eqref{eq:GGS}?
A possible answer to this question is provided from the point of view of dynamical typicality, or canonical typicality (see refs.~\cite{Goldstein2006PRL_canonical,Popescu2006NPhys_entanglement,Linden2009_evolution}).  There, the idea is that on a system and a reservoir, any state chosen at random in a microcanonical subspace has a reduced state on the single system that looks thermal with overwhelming probability.  If the evolution is sufficiently ergodic, exploring the full accessible state space, then the system will look thermal over an overwhelmingly large fraction of time.

The microcanonical approach provides a useful tool to analyze the situation of noncommuting multiple charges in the context of dynamical typicality~\cite{MicroCanonical}.
Consider as above $N$ copies of a system $s$ with a collection of physical quantities $\{ \bar{Q}_j \}$.  Here we assume that all charges commute with the Hamiltonian $H$ governing the time evolution ($H$ may or may not be included in the collection $\{ Q_j \}$).
Assume that the $N$ systems are in a state $\ket\psi$ which lives in an approximate microcanonical subspace $\mathcal{M}$ corresponding to charge values $\{ v_j \}$.  Canonical typicality asserts that if $\ket\psi$ is chosen uniformly at random in the subspace, then on average the reduced state on any single system $\ell$ is well approximated by the reduced microcanonical state:
\begin{align}
  \left\langle 
  \frac12\,\norm[\Big]{ \tr_{1,\ldots\ell-1,\ell+1\ldots,N}(\psi) - \tr_{1,\ldots\ell-1,\ell+1\ldots,N}(\Omega) }_1
  \right\rangle_{\psi}
  \leqslant \frac{\dim(s)}{\sqrt{\dim(\mathcal{M})}}\ ,
  \label{eq:canonical-typicality-reduced-avg-close-to-GGS}
\end{align}
noting that $\dim(s)/\sqrt{\dim(\mathcal{M})}\to 0$ as $N\to\infty$ because $\dim(\mathcal{M})$ scales exponentially in $N$ while $\dim(s)$ is constant.  Combined with~\eqref{eq:approx-microcan-gives-reduced-thermal-state}, this tells us that, with high probability, the reduced state of $\ket\psi\in\mathcal{M}$ on a single system is close to the generalized Gibbs state.

Under suitable assumptions, the evolution of the system is ergodic, meaning that~\eqref{eq:canonical-typicality-reduced-avg-close-to-GGS} holds as an average over the time evolution~\cite{Linden2009_evolution}.  More precisely, for almost all initial states $\ket{\psi(0)}$, and denoting by $\ket{\psi(t)}$ the corresponding time-evolved state, we have
\begin{align}
  \lim_{T\to\infty}\frac1T\int_0^T
  \frac12\,\norm[\Big]{ \tr_{1,\ldots\ell-1,\ell+1\ldots,N}(\psi(t)) - \tr_{1,\ldots\ell-1,\ell+1\ldots,N}(\Omega) }_1
  \, dt\ 
  \leqslant\  \frac{\dim(s)}{\sqrt{\dim(\mathcal{M})}}\ .
\end{align}
Combining this with~\eqref{eq:approx-microcan-gives-reduced-thermal-state}, we see that for almost all initial states $\ket{\psi(0)}\in\mathcal{M}$, over time the state on a single system stays close to the generalized Gibbs state.

This treatment only scratches the surface of the question of equilibration in the presence of multiple conserved quantities, and a more detailed analysis is still an open question.






\section{Resource theory approach}
In this section we generalise the resource theoretic framework introduced in \cite{SecondLawsChap} to the framework of multiple conserved quantities. We focus, in particular, on single-shot thermodynamics and work extraction in this paradigm. This approach that was first put on firm footing in \cite{janzing2000thermodynamic,horodecki2013fundamental, Brandao2013,Brandao2015PNAS_secondlaws}  and provided a single-shot and resource theoretic approach for one conserved quantity -- energy. The motivation to extend these ideas to more quantities lies in the desire to understand the privileged status that energy has in our world. What thermodynamics can we do when energy plays no important role and all conserved quantities appear on a level playing field? Does this framework give us access to new physics and how does it deviate from what we observe?

We proceed to tell the story in the standard general framework of thermodynamics, that consists of a thermal bath $b$, an out-of-equilibrium system $s$ in a state $\rho_s$, and a number of batteries where we can store the extracted resources. 
Recall that when building a resource theory one must first fix the state space and the allowed state transformations. Following this, one investigates the resulting structure on the state space and the properties arising from such assignments specifying $(i)$ the class of \textit{free operations}, that can be applied at no cost; $(ii)$ the class of \textit{free states} that can be prepared at no cost, i.e. the states that are invariant under the class of free operations.

When the resource theory concerns thermodynamics with one conserved quantity, the free operations are generated by energy-preserving unitaries, $[U, H] = 0$, where $H = H_s + H_b$ is additive over the system and bath. The free states are the familiar Gibbs states $\tau = e^{-\beta H_s}/\mathcal{Z}$.
 When the system has not one, but two conserved quantities $Q_1 = A_s$ and $Q_2 = B_s$, the general recipe for constructing the resource theory does not change.  We take the state space to be the joint Hilbert space of the system and bath, and specify the allowed transformations to be global unitaries which preserve the additive quantities $A$ as well as $B$, namely $[U, A] = [U, B] = 0$. The free states become the \textit{generalised} Gibbs states $\tau = e^{-(\beta_A A_s + \beta_B B_s)}/\mathcal{Z}$. 
The relationship between $A$ and $B$ is important and affects the results, depending on whether they are functionally related or how they commute.

\subsection{The Generalised Gibbs State}

We will begin by stating the generalised thermal state  -- the free state of our theory -- and then proceed to derive it from resource-theoretic considerations. To this end we follow Jaynes \cite{Jaynes1, Jaynes2} and take a `thermal bath' to be a collection of particles each in the generalised thermal state given in \eqref{eq:GGS}.

 At this moment, we place no restrictions on the conserved quantities -- they may or may not commute and they may or may not be functionally related, moreover energy need not even be one of these quantities. 	
To derive this state, we return to standard thermodynamics and recall that when restricting considerations to energy, there are two ways to define
the thermal state: either by maximising the von Neumann entropy subject to the average energy being given \textit{or} by minimising the free energy given the inverse temperature
\begin{align}\label{eq:statestan}
\tau(\beta_E) = \frac{e^{-\beta_E E}}{\mathcal{Z}}
\qquad\quad \begin{cases}
	& \text{maximises } \;\;S(\rho) = -\tr[\rho \log \rho] \quad \quad\;\text{given} \quad  \overline{ E}\\
    & \text{minimises } \;\;F(\rho) = \langle E\rangle_\rho  - T_ES(\rho)\quad \text{given} \quad \beta_E
  \end{cases}
\end{align}
where $T_E = 1/\beta_E$ and the average $\tr[\rho E] = \overline{E}$. The thermal state is the state which simultaneously extremises two quantities -- the entropy and the free energy. In similar spirit, we would like for the generalised thermal state in Eq.~\eqref{eq:GGS} to be the state that also extremises two functions. From Jaynes we know that the generalised thermal state is still the state that maximises the entropy given the expectation values. The second quantity we are looking for is then the one that is minimised by the generalised thermal state. Previously, this was the free energy where the constant multiplying the entropy was the temperature. We do not have a notion of multiple entropies to couple the temperatures in order to extend the definition, but we can couple the inverse temperatures to the conserved quantities. We define the \textit{free entropy}
\begin{definition}[Free entropy]
The free entropy of a system $\rho$ is a map from density matrices to real numbers $\tilde{F}: \mathcal{S}(\mathcal{H})\longrightarrow \mathbb{R}$,
\begin{align}\label{eq:freeent}
\tilde{F}(\rho) = \sum_i \beta_i \langle Q_i \rangle_\rho- S(\rho)
\,.
\end{align}
where  $\langle Q_i \rangle_\rho = \tr[Q_i\rho]$ and $S$ is the von Neumann entropy. 
\end{definition}

The generalised thermal state is then the state that simultaneously maximises the von Neumann entropy and minimises the free entropy.
\begin{theorem}
  The generalised thermal state
\begin{align}\label{eq:stategen}
\tau(\mathbold{\beta}) = \frac{e^{-\sum_i \beta_i Q_i}}{\mathcal{Z}}
\qquad\quad \begin{cases}
	& \text{maximises } \;\;S(\rho) = -\tr[\rho \log \rho] \quad\quad\;\quad \text{given}  \quad  \overline{Q_1}  , \cdots , \overline{ Q_k}\\
    & \text{minimises } \;\;\tilde{F}(\rho) = \sum_i \beta_i \langle Q_i\rangle - S(\rho)\quad \text{given} \quad  \beta_1, \cdots , \beta_k
  \end{cases}
\end{align}
where we have collected the inverse temperatures into a vector $\mathbold{\beta} = (\beta_1, \cdots, \beta_k)$ and $\mathcal{Z} = \tr[e^{-\sum_i \beta_i Q_i}]$.
\end{theorem}
\noindent The thermal state is diagonal in the basis of the large observable $R = \sum_i \beta_i Q_i$. If the eigenvalues of $R$ are $\{r_i\}$ then the probability of a particle being in the $i-$th state is $p_i = e^{-r_i}/\mathcal{Z}$. If the observables $Q_i$ commute then the probabilities becomes $p_i = e^{-(\beta_1 q^1_i +\cdots +\beta_k q^k_i )}/\mathcal{Z}$, where $q^k_i$ is the $i-$th eigenvalue of observable $k$. 
 Eq.~\eqref{eq:stategen} can also be viewed as a duality relation for an optimization problem. The free entropy is a dimensionless function of state and can be viewed as the dual quantity to entropy. While the entropy is a function of extensive quantities $Q_i$ and is maximised, performing the Legendre transform yields the dual function $\tilde{F}$ of intensive quantities $\beta_i$ which is minimized. 
The proofs of Eq.~\eqref{eq:stategen} for both commuting and non-commuting quantum observables can be found in \cite{MultipleConserved}. Proofs from a Bayesian perspective were first presented by Jaynes \cite{Jaynes1} and via a different method more recently by Liu \cite{Kai2006}.

\begin{example}[Energy and angular momentum] 
 An example of two commuting (and functionally dependent) conserved observables are energy $E$ (i.e. the Hamiltonian) and angular momentum $L$,  where $E = L^2/2I$ and $I$ is the moment of inertia. A thermal bath characterised by energy and angular momentum is a collection of thermal states of the form $\tau(\beta_E, \beta_L) = e^{-(\beta_E E +\beta_L L)}/\mathcal{Z}$. One may picture a `sea' of flywheels, each with different moments of inertia, rotating with different angular momenta (clockwise/anticlockwise at different rates). Picking a particular flywheel from the bath, the likelihood that it has angular momentum $L_i$ and energy $E_i =L_i^2/2I$  is given by the probability $p_i = e^{-(\beta_E E_i + \beta_L L_i)} /\mathcal{Z}$.
\end{example}

There are two arguments to encourage the reader to adopt the new `free entropy' definition. In the standard definition in Eq.~\eqref{eq:statestan}, the free energy was minimised only for positive temperature $T_E>0$ (and consequently positive inverse temperature $\beta_E>0$). For negative temperature\footnote{In a finite quantised spectrum, a population inversion with a Gibbs profile is equivalent to a negative temperature system. Since population inversion and thus negative temperatures are accessible in experimental settings, it is preferable to have an argument that is independent of the sign of the temperature.}, one must reverse the argument to conclude that the thermal state $\tau(\beta_E)$ \textit{maximises} the free energy. On the other hand, the generalised thermal state in \eqref{eq:stategen} is the state that minimises the free entropy for all inverse temperatures $\beta_i$ regardless of whether they are positive or negative, lending a kind of universality to the new definition and standing up to arguments that the free entropy is a quantity that is purely `rescaled'. The second observation is that since each term $\beta_i Q_i$ is dimensionless, they can be regarded as entropy-like quantities, not least because the difference in free entropy between any state $\rho$ and the generalised thermal state is equal to the relative entropy difference between those two states. 
\begin{align}
 \tilde{F}(\rho_s)- \tilde{F}(\tau(\mathbold{\beta}))  = \; \Delta \tilde{F} = D(\rho_s\|\tau (\mathbold{\beta}))
\,.
\end{align} 
where $D(\rho\|\sigma) = - S(\rho)-\tr[\rho \log \sigma].$ In this way, all quantities are placed on equal footing and energy (should it appear as one of the conserved quantities) plays no special role.

\subsection{Second laws}
We will look to derive the second law from operational principles. For clarity, and to avoid clutter we will concentrate on two quantities $A$ and $B$, since the generalisation to $k$ quantities follows naturally. In the standard picture of thermodynamics in the single-shot regime one takes a finite number of bath states and a quantum system, initially uncorrelated and out-of-equilibrium with respect to the bath $\rho_s \otimes \tau(\beta_E)^{\otimes N}$ and looks for energy conserving unitaries which `do work',  for instance to raise a weight. To this end, one effects the transformation 
$U_{sb}(\,\rho_s \otimes \tau(\beta_E)^{\otimes N} )\,U_{sb}^\dagger$. The amount of energetic work $\Delta W_E$ that can be extracted from the system, i.e. the amount by which an arbitrary weight (in any level) can be raised is constrained by the second law 
\begin{align}
 \Delta W_E \le - \Delta F_s 
\,.
\end{align} 
where $F$ is the free energy of the system as given in \eqref{eq:statestan}. Equality is achieved when the right unitary is executed and the system thermalises to be indistinguishable from the states of the bath $\rho_s \rightarrow \tau(\beta_E)$. A Kelvin-Planck type statement of the second law is that `there is no way to extract energy from a single thermal bath'.  In the picture involving two quantities $A$ and $B$, the recipe is much the same. One starts by taking a finite number of systems from a generalised thermal bath and an out-of-equilibrium quantum system, which is initially uncorrelated from the bath $\rho_s \otimes \tau({\beta_A, \beta_B})^{\otimes N}$. The focus is now on extracting $A$ and $B$ from the system and storing them in their associated batteries. As the quantities are additive, the amount in the system and bath can be treated independently. The $A$- and $B$-type work are defined in such a way that they automatically include the \textit{first law of thermodynamics}
\begin{align}\label{eq:firstlaw}
\Delta W_A = -\Delta A_s -\Delta A_b \qquad \quad
\Delta W_B= -\Delta B_s -\Delta B_b 
\end{align}
where $\Delta A_s  = \tr[A_s (\rho_s' -\rho_s)] $ and  $\Delta A_b  = \tr[A_b (\rho_b' -\rho_b)] $ and analogously for $B$. When talking about work extraction there are two ways to proceed: either by including batteries implicitly, or explicitly, in the formalism.\\

\noindent\textbf{Implicit battery.} The global amount of $A$ and $B$ in the system and bath change. These changes are defined as `A-type' work $\Delta W_A$  and `B-type' work $\Delta W_B$, which are quantities that have been extracted from (or done on) the global system. Due to the conservation laws \eqref{eq:firstlaw}, when  $A$ or $B$ of the system and bath change, this change is compensated by a corresponding change to the external environment, i.e. the implicit battery. The transformations one considers are of the form $U_{sb}(\,\rho_s \otimes \tau(\beta_A, \beta_B)^{\otimes N} )\,U_{sb}^\dagger$.\\

\noindent\textbf{Explicit battery}. An explicit battery is a mathematical model for a work storage device. It accepts only a single type of work (i.e. an `A-type' battery will only accept `A-type' work, and a `B-type' battery only `B-type' work). The transformations one considers are of the form $U_{}(\,\rho_s \otimes \tau(\beta_a, \beta_B)^{\otimes N} \otimes \rho_{w_A} \otimes \rho_{w_B} )\,U_{}^\dagger$, where $\rho_{w_A}$ is the state of battery-$A$, or `weight'-$A$ and similarly for $\rho_{w_B}$. The global $U$ acts on all four systems. 
\begin{theorem}[The Second Law]\label{theo:secondlaw}
Given a generalised thermal state characterised by two inverse temperatures $\tau(\beta_A, \beta_B)$ and an out of equilibrium quantum system $\rho_s$ the maximum amount of $A$- and $B$-type work which one can extract is constrained in the following way
\begin{align}
 \beta_A \Delta W_A + \beta_B \Delta W_B \le - \Delta \tilde{F}_s  \label{eq:secondlaw}
\,.\
\end{align} 
\end{theorem}

\begin{corollary}[The Second Law with only a bath]\label{corr:1}
Given a generalised thermal bath characterised by two inverse temperatures $\tau(\beta_A, \beta_B)$ (and therefore a quantum system which is also thermal $\rho_s = \tau(\beta_A, \beta_B)$) the maximum amount of $A$- and $B$-type work which one can extract is constrained in the following way
\begin{align}\label{eq:corr}
 \beta_A \Delta W_A + \beta_B \Delta W_B \le 0
\,.
\end{align} 
\end{corollary}
\noindent

\begin{proof}[Proof of the Second Law: implicit batteries] (See also \cite{UzdinChap})
 First, since we restrict to unitary transformations the total entropy of the global system remains unchanged, $S(\rho'_{sb}) = S(\rho_{sb})$, and from the fact that the system and bath are initially uncorrelated, along with sub-additivity of von Neumann entropy, we have
\begin{align}\label{eq:entropy}
S(\rho_s) + S(\rho_b)= S(\rho_{sb}) \qquad &\text{and}\qquad
S(\rho_s^\prime) + S(\rho_b^\prime)\ge S(\rho_{sb}^\prime)\\
\Rightarrow \Delta S_s + &\Delta S_b \geq \Delta S_{sb} = 0 \label{zeroent}
\end{align}
where $\Delta S_s = S(\rho'_s) - S(\rho_s)$, and analogously for $\Delta S_b$ and $\Delta S_{sb}$. Second, since the bath starts in the thermal state $\tau(\beta_A,\beta_B)$, which is in a minimum of the free entropy by definition, its free entropy can only increase, thus
\begin{equation}\label{eq:repeat1}
\Delta \tilde{F}_b = \beta_A \Delta A_b + \beta_B \Delta B_b - \Delta S_b \geq 0
\end{equation}
Now, eliminating all quantities on the bath, by substituting from the first laws \eqref{eq:firstlaw} and from \eqref{zeroent}, we finally arrive at 
\begin{align}\label{eq:repeat2}
-\beta_A (\Delta A_s + \Delta W_A) - \beta_B (\Delta B_s + \Delta W_B) + \Delta S_s &\geq 0\\
 \beta_A \Delta W_A + \beta_B \Delta W_B &\leq -\Delta \tilde{F}_s. 
\end{align}
\end{proof}
\noindent Note that the proof does not rely on any particular properties of $A$ and $B$, which need not even commute, and so in the implicit battery case, the result is trivially generalised to $k$ conserved quantities. 
\begin{proof}[Proof sketch of the second law: explict batteries.] In order to prove the second law with explicit batteries, one must supply a model.  The batteries are modelled as `weights' living on ladders such that the value of observable $A$ on the battery is proportional to the position operator $A_{w_A} =c_a\hat{x}_a$, where $c_a$ is a constant of appropriate units. The corresponding translation operator $\Gamma^{\epsilon}_{\omega_A} = \exp (-i\epsilon \hat{p}_a)$
effects the  transformation
$\Gamma^{\epsilon}_{\omega_A} \ket{x_a} = \ket{x+\epsilon}$ on unnormalised position states of the pointer (where $\hat{p_a}$ is the momentum operator, canonically conjugate to the position).  The objective is to have a reliable `meter' for a system of any dimension and the fact that the ladders are unbounded allows one to avoid boundary details, which are not the primary focus of this discussion\footnote{One can imagine a ladder that is very high (orders of magnitude greater than the number of system levels). If one is in the centre of this ladder then for all practical purposes it is unbounded from above and below. Similarly, if the initial systems are $d$-dimensional then one can also use a finite ladder and avoid boundary effects by initialising it in a state whose support is greater than $d$-levels about the ground state.}.
Work is defined as $\Delta W_A = \Delta A_{w_A}$, and analogously for $B$.  In order to eliminate the possibility of cheating one must introduce conditions on the batteries. ‘Cheating’ can encompass a variety of actions, such as bringing in ancillas, which would contribute as external sources of free entropy or work  by using the batteries as cold reservoirs to embezzle work. The four assumptions that are made are:
\begin{enumerate}
\itemsep0em 
\item \label{rule:1}\textit{Independence}: Batteries are independent of one another and only accept and store one type of conserved quantity. Each quantity is assigned its own battery system.
\item \textit{First laws:} The set of allowed operations are global unitaries $U$ on the bath, system and batteries that conserve $A$ and $B$, in other words  $[U, A_b + A_s + A_{w_A}] = [U, B_b + B_s + B_{w_B}] = 0 $, which impose the first laws of thermodynamics.\footnote{Unitaries are chosen as opposed to more general completely positive (CP) maps in order not to use external ancillas in non-thermal states as sources of work.}
\item \textit{Translational invariance: }Only displacements in the position on the ladders are important. This implies all $U$ should commute with translation operators $\Gamma^{\epsilon}_{\omega_{A}}, \Gamma^{\epsilon}_{\omega_{B}}, \cdots $ on each battery.
\item \textit{Initially uncorrelated: }The initial state of the systems is of the product form $\rho_s\otimes \tau^{\otimes N}\otimes \rho_{w_A}\otimes \rho_{w_B}$.
\end{enumerate}
One then uses a theorem (see \cite{MultipleConserved}) that says that unitary evolutions $U$ which are weight-translation invariant cannot decrease the entropy of the system and bath
$
\Delta S(\rho_{sb})\ge 0
$. This then allows one to say that $\Delta S_b \ge -\Delta S_s$ and repeat the arguments as in the implicit battery case in Eqns.~\eqref{eq:repeat1} --\eqref{eq:repeat2} to arrive at the second law. 
\end{proof}
\begin{example}[Energy and angular momentum.]
The battery used to store energy can be thought of as a weight and the battery storing the angular momentum a flywheel. We wish for the batteries to operate independently, as per assumption \eqref{rule:1}. The energy of the weight is simply the Hamiltonian. If we take the flywheel to be very massive, i.e. the momentum of inertia very large, then the energy of the flywheel becomes essentially independent of its rotational property $E_{fly} = \frac{L^2}{2I}\approx 0$. Varying the angular momentum changes $E_{fly}$ very little, thus the two batteries can effectively operate independently of one another. 
\end{example}

\subsection{Applications}
\subsubsection{Extracting Resources}
In \cite{MultipleConserved} it was shown that for commuting quantities $[A, B] = 0$, the second law \eqref{eq:secondlaw} is \textit{tight}.  In order to demonstrate tightness one needs to provide an explicit unitary $U$. Let $p_i$ represent the populations of the system, i.e., the probability of the system to occupy the $i$-th joint eigenstate of $A$ and $B$. The effect of the unitary is to create a small change $\delta p$ between two of the levels $p_0' = p_0 +\delta p$ and $p_1' = p_1 - \delta_p$ of the system. One must also show that the change in free entropy of the bath due to this $U$ is small
$
\Delta\tilde{F}_b= O(\delta p^2),
$
as well as the fact that the system and bath remain effectively uncorrelated after the interaction $\Delta S_s +\Delta S_b =  O(\delta p^2)$.
\begin{equation}
\beta_A \Delta W_A + \beta_B \Delta W_B = -\Delta \tilde{F}_s + O(\delta p^2)
\end{equation}
i.e.\@ that up to a correction of order $O(\delta p^2)$, the combination of conserved quantities extracted, which themselves are of order $O({\delta p})$, matches the change in free entropy of the system. The state of the system is now a little closer to thermal. Thus, by repeating the process $O(\frac{1}{\delta p})$ times (each time taking $N$ new bath states) one can implement a protocol which transforms $\rho_s \to \tau_s(\beta_A,\beta_B)$, whereby in each stage the population changes between two states by order $O(\delta p)$, and such that in the end
\begin{equation}
\beta_A \Delta W_A + \beta_B \Delta W_B = -\Delta \tilde{F}_s + O(\delta p).
\end{equation}
Thus, by taking $\delta p$ sufficiently small one can approach reversibility. In the same work the authors present a variant of the protocol which is \textit{robust}. Robust, in this case means that should an experimenter have access to a generalised thermal bath $\tau(\beta_A, \beta_A)$, but they can only characterise the inverse temperatures to some finite precision, or `window' $x_1\le\beta_A\le x_2$ and $y_1\le\beta_B\le y_2$, then there still exists a procedure which will saturate the second law\footnote{Up to a measure zero set of inverse temperatures which present `pathological' cases.}. These results are valid for commuting quantities $A$ and $B$, and can be easily generalised to $k$ quantities. For the case that they do not commute $[A, B]\neq 0$ there is no known protocol to extract work from a quantum system, and it remains an open question to show whether or not it is possible to saturate the second law. 
\subsubsection{Trading Resources}
Using Corollary \ref{corr:1} one can arrive at one of the central results of this formalism, namely, that conserved quantities can be \textit{traded.} In particular, for commuting quantities $[A, B]=0$ in the presence of a single bath one can show that there exists a unitary acting solely on the bath $U_b$ (given in \cite{MultipleConserved}) such that  $U_b \,\tau(\beta_A, \beta_B)\,U^\dagger_b$ results 
in changes in $\Delta A_b$ and $\Delta B_b$ that are very large compared with the change in free entropy. 
\begin{align}\label{eq:trade}
\Delta\tilde{F}_b =  \beta_A \Delta A_b + \beta_B \Delta B_b \approx
 \epsilon 
\,
\quad
\Rightarrow 
\quad 	
\Delta A_b  \approx -  \frac{\beta_B}{\beta_A} \Delta_b B +\frac{\epsilon}{\beta_A}
\,.
\end{align}
When $|\tfrac{\Delta A}
{\Delta \tilde{F}_b}|$ and $|\tfrac{\Delta B}
{\Delta \tilde{F}_b} |$ are sufficiently large and $\epsilon$ is sufficiently small, we say that these quantities can be trade \textit{reversibly}. The exchange rate for this trade is given by the ratio ${\beta_B}/{\beta_A}$.\footnote{Note that although $U_b$ does not change the free entropy of the bath, it \textit{does} change the state of the bath, such that the state after the interaction could be very far from $\tau(\beta_A, \beta_B)$} 
Note that it is important that $\Delta \tilde{F}_b \neq 0$ --- since the thermal state is the unique state which minimises $\tilde{F}_b$, this would require that the bath is left completely unchanged, resulting in trivial changes $\Delta A_b = \Delta B_b = 0. $ This is in stark contrast to the standard picture, where the thermal state is useless for all practical purposes. Here, the analogous Kelvin-Planck statement of the second law becomes becomes `there is no way to extract the linear combination $\beta_A A+\beta_B B$ from a single, generalised thermal bath'. This demonstrates the interconversion of resources, which is illustrated in the following example. 
\begin{example}[Energy and angular momentum]
Imagine that a protocol has successfully effected the transformation $\rho_s\otimes \tau(\beta_E, \beta_L)^{\otimes N} \rightarrow \tau(\beta_E, \beta_L)\otimes \tau(\beta_E, \beta_L)^{\otimes N} $.  The free entropy of the system is spent and it has now become thermal, an extra particle in the bath. Both energy and angular momentum have been extracted and consequently the weight has been raised to a certain height and the flywheel is spinning with some angular momentum.  Eq.~\eqref{eq:trade} tells us that we can raise the weight even higher at the expense of slowing down the fly-wheel (or vice versa).
The currency for this trade is given by the ratio $\beta_E/\beta_L$. 
\end{example}
\noindent The existence and construction of unitary for trading quantities which do not commute $[A, B]\neq 0$ remains an open problem. The results discussed in this section, as well as extensions are summarised in the table below.
\begin{table}[h!]
\begin{tabular}{cc|c|c}
                                                                                                                 &          & Commuting & Non-commuting \\ \hline
\multicolumn{1}{l|}{\multirow{2}{*}{\begin{tabular}[c]{@{}l@{}}Implicit\\ batteries\end{tabular}}}                  & 2nd law  & $\checkmark $         & $\checkmark $             \\ \cline{2-4} 
\multicolumn{1}{l|}{}                                                                                               & protocol & $\checkmark $         & $\checkmark $             \\ \hline
\multicolumn{1}{l|}{\multirow{2}{*}{\begin{tabular}[c]{@{}l@{}}Explicit batteries\\ (strict cons.)\end{tabular}}} & 2nd law  & $\checkmark $         & $\checkmark $             \\ \cline{2-4} 
\multicolumn{1}{l|}{}                                                                                               & protocol & $\checkmark $         & ?             \\ \hline
\multicolumn{1}{l|}{\multirow{2}{*}{\begin{tabular}[c]{@{}l@{}}Explicit batteries \\ (ave. cons.)\end{tabular}}}   & 2nd law  & $\checkmark $         & $\checkmark $            \\ \cline{2-4} 
\multicolumn{1}{l|}{}                                                                                               & protocol & $\checkmark $         & $\checkmark $*            
\end{tabular}
\caption{\label{table1} Summary of the results for the 'resource theory' with multiple conserved observables. Here, `protocol' means the existence and construction of a unitary transformation that demonstrates extraction and trade of resources. Explicit battery models can be split into those which conserve observables strictly and those which conserve them on average. * indicates that a unitary has been found for explicit batteries that have continuous (i.e.\@ not discrete) spectra.  }
\end{table}


\subsubsection{Landauer with multiple charges}

The topics of Landauer erasure \cite{landauer61}, and the Szilard engine \cite{Szilard1929} illustrate subtle connections between information and thermodynamics. The most basic concept in information is that of a `bit', which is simply the answer to some `yes/no' question. In the case of quantum mechanics, these yes/no states are labeled as $|0\>$ and $|1\>$. Such a system contains a single bit of information if it is in the (maximally) mixed state $\rho_1 = \frac{1}{2} |0\>\<0| + \frac{1}{2}|1\>\<1|$. 

Landauer showed that logically irreversible processing, such as resetting a single bit its $|0\>$ state, necessarily incurs a corresponding entropy increase of the environment, which in turn causes the dissipation of $kT\ln(2)$ heat~\cite{landauer61}.  This energy must be provided in the form of work; this observation provided fundamental insights to the thermodynamics of computing, and was instrumental to the exorcism of Maxwell's demon (see ~\cite{Bennett1982IJTP_ThermodynOfComp,Bennett2003_NotesLP,MaxwellDemonChap}  for recent Maxwell demon experiments).
It is therefore a natural question to ask whether this thermodynamic cost can be furnished in terms of other physical quantities:
How does Landauer erasure behave in the context of multiple conserved quantities? Do quantum-mechanical effects, such as non-commutativity, affect this erasure?
To address this we formulate erasure with multiple charges, and present a generalized Landauer result. This topic is discussed in more detail in \cite{InformationErasureChap}, and in the works \cite{vaccaro2011information,Varcoe,NonCommutativity,croucher2017discrete}. Here we present a simple entropic account of erasure.

As before, we allow for an arbitrary number of conserved charges $\{Q_1, \dots, Q_k\}$, and make use of the generalized Gibbs state
\begin{equation}
\tau_B = \frac{1}{\mathcal{Z}} e^{-\sum_i \beta_i Q_i}.
\end{equation}
Then a general erasure procedure can be described by the unitary transformation $\rho_S \otimes \tau_B \mapsto \rho'_{SB}:=U(\rho_S \otimes \tau_B)U^\dagger $ for some unitary $U$ acting on both systems. Now we can consider the mutual information between $S$ and $B$, which is given by $I(S:B) := S(\rho'_A) + S(\rho'_B) - S(\rho'_{SB})$, where $\rho'_S = \tr_B [ \rho'_{SB}]$ and similarly for $\rho'_B$.

Since the von Neumann entropy is unitarily invariant we have that $S(\rho'_{SB}) = S(\rho_S \otimes \tau_B) = S(\rho_S) + S(\tau_B)$ and so 
\begin{equation}\label{diff}
\Delta S_S +  \Delta S_B  = I(S:B).
\end{equation}
We now exploit the basic features of the generalized Gibbs state. It is readily seen that
\begin{equation}\label{DeltaSB}
\Delta S_B  = \sum_i \beta_i \tr [ Q_i (\rho'_B - \tau_B)] - D(\rho'_B ||\tau_B),
\end{equation}
where $D(\rho||\sigma)=\tr [ \rho \log \rho - \rho \log \sigma]$ is the relative entropy. We now assume that the generalized thermal bath is sufficiently large that it is unaffected in the thermodynamic process, and so $\rho'_B \approx \tau_B$ which means the last term in equation (\ref{DeltaSB}) can be neglected. We now define a $Q_i$ heat flow via $\<Q_i\> :=  \tr [ Q_i (\rho'_B - \tau_B)] $. Inserting these into equation (\ref{diff}) gives us that
\begin{equation}
\sum_i \beta_i \<Q_i\> = -\Delta S_S + I(S:B).
\end{equation}
However the mutual information is always non-negative and so we deduce that $\sum_k \beta_k \<Q_k\> \ge -\Delta S_S$. This is the generalized Landauer bound in the presence of arbitrary many conserved charges and where the erasure need not be perfect. For the case of complete erasure we have an initial state $\rho_S$ with entropy $\ln 2$ (a single bit) that is sent to some default state with zero entropy. Thus for complete erasure we have the general erasure bound,   
\begin{equation}
\sum_i \beta_i \<Q_i\> \ge \ln 2.
\end{equation}
This relation again highlights the ability to trade the difference resources on an equal footing within a thermodynamic process. A key thing to notice however, is that nowhere did we use the commutation relations between the charges. In other words, the above Landauer bound applies equally well for commuting and non-commuting cases.


\section{Outlook}

A key aspect of quantum mechanics is the non-commutativity of observables (such as position and momentum), and the resultant complementarity which prohibits assigning definite values simultaneously to these observables. In this chapter, we have shown how multiple quantum observables that may possess non-trivial commutation relations, may be incorporated into a framework of quantum thermodynamics. Subtleties aside, non-commutativity does not create an obstacle to the generalized Gibbs ensemble playing the same role as in classical statistical mechanics.

That said, a key goal of the recent quantum thermodynamics program is precisely to identify quantum-mechanical signatures that signal a departure from classical physics. In this regard one would expect that non-commutativity should produce such a thermodynamic signature, however the analysis that has so far been under-taken has not revealed any. For example, the form for the generalized Landauer erasure involving multiple conserved charges $\{Q_i\}$ does not differ if these charges commute or not. However there exist a range of reasons why one expects such a signature. The MaxEnt procedure \cite{Jaynes1,Jaynes2} singles out the generalized Gibbs state subject to constraints on the expectation value of the microscopically conserved observables $\{\<Q_1\>, \<Q_2\>, \<Q_3\>,\dots \}$. However it is known that the MaxEnt procedure possesses a \emph{discontinuity} in the quantum-mechanical case: arbitrarily small changes in values of the $\<Q_i\>$ constraints can produce large changes in the associated generalized Gibbs state. This feature is due to non-commutativity \cite{weis2012entropy, barndorff2014information}, and has connections with quantum phase transitions in many-body quantum systems \cite{chen2015discontinuity}. In contrast to the analysis conducted here, the way in which this non-commutativity is detected is through the varying of the external constraints (for example the switching of classical field strengths). Recent quantum information-theoretic approaches are very good at explicitly accounting for non-classical resources (such as quantum coherence), however this strength becomes a weakness when it comes to varying the constraining fields. This is because to do such a change necessarily involves a non-trivial use of coherence that must be accounted for thermodynamically. Such an analysis has been done in the case of an effective change in the energy eigenbasis \cite{korzekwa2016extraction}, and so a similar approach could be taken for the case of multiple conserved observables. 

Recent work has formulated a trade-off relation for energy and time in quantum thermodynamics \cite{kwon2017clock}, and a natural direction would be to extend such a relation to include non-commuting conserved charges. An alternative way to detecting the effects of non-commutativity would be to exploit the fact that the microscopically conserved charges have an associated symmetry group that explicitly depends on their commutation relations -- therefore the structure of the state interconversions, admissible within the thermodynamic framework, must carry a signature of any non-commutativity.  Also, recently a complete set of conditions for state interconversions for such a general scenario have been derived \cite{gour2017quantum}, however a detailed study of these conditions still remains to be done. 

One other direction that deserves exploration is the approximate micro-canonical subspaces \cite{MicroCanonical} and their precise relation to typicality in the case of non-commuting charges.



\bigskip
\acknowledgements
ACKNOWLEDGEMENTS

EHM is funded by the EPSRC, DJ is supported by the Royal Society. YG acknowledges funding from the FWF START grant Y879-N27. PhF acknowledges support from the SNSF through the Early PostDoc. Mobility Fellowship No.~{P2EZP2\_165239} hosted by the Institute for Quantum Information and Matter
(IQIM) at Caltech, as well as from the National Science Foundation.



\bibliography{Multiple_References_1}

\end{document}